\documentstyle[epsfig]{article}

\def\1ad{\mbox{\normalsize $^1$}}
\def\2ad{\mbox{\normalsize $^2$}}
\def\3ad{\mbox{\normalsize $^3$}}
\def\4ad{\mbox{\normalsize $^4$}}
\def\5ad{\mbox{\normalsize $^5$}}
\def\6ad{\mbox{\normalsize $^6$}}
\def\7ad{\mbox{\normalsize $^7$}}
\def\8ad{\mbox{\normalsize $^8$}}
\def\makefront{\vspace*{1cm}\begin{center}
\def\newtitleline{\\ \vskip 5pt}
{\Large\bf\titleline}\\
\vskip 1truecm
{\large\bf\authors}\\
\vskip 5truemm
\addresses
\end{center}
\vskip 1truecm
{\bf Abstract:}
\abstracttext
\vskip 1truecm}
\begin {document}
\def\titleline{Interference of spontaneously emitted photons}
\def\authors{Almut Beige, Christian Sch\"on and Jiannis Pachos}
\def\addresses{Max-Planck-Institut f\"ur Quantenoptik, 85748 Garching,
Germany \\ {\tt e-mail: a.beige@mpq.mpg.de}} \def\abstracttext{ We
discuss an experimental setup where two laser-driven atoms
spontaneously emit photons and every photon causes a
``click'' at a point on a screen. By deriving the probability density
for an emission into a certain direction from basic quantum
mechanical principles we predict a spatial interference
pattern. Similarities and differences with the classical double-slit
experiment are discussed.} \large \makefront

\section{Introduction}

In 1930 Werner Heisenberg wrote in \cite{Heisenberg}, ``It is very difficult 
for us to conceive the fact that the theory of photons does not conflict with
the requirements of the Maxwell equations. There have been attempts to
avoid the contradiction by finding solutions of the latter which
represent `needle' radiation (unidirectional beams), but the results
could not be satisfactorily interpreted until the principles of the
quantum theory had been elucidated. These show us that whenever an
experiment is capable of furnishing information regarding the
direction of emission of a photon, its results are precisely those
which would be predicted from a solution of the Maxwell equations of
the needle type (...).'' 

\vspace*{-0.5cm}
\noindent \begin{figure}[h]
\begin{center}
\epsfig{file=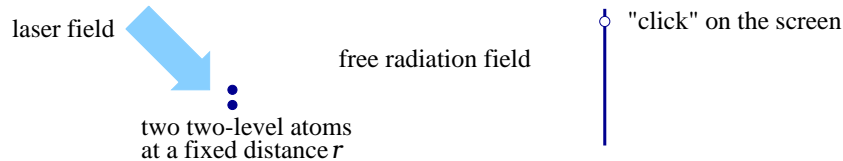,width=11cm}\\[0.1cm]
\caption{Experimental setup. Two two-level atoms are placed at a fixed
distance $r$ from each other. Both are coupled to the same free
radiation field and are continuously driven by a resonant laser. This
leads to spontaneous photon emissions. Each photon causes a ``click"
at a point on a screen.} \label{fig1}
\end{center}
\end{figure}

\vspace*{-0.5cm} A simple quantum mechanical experiment that combines
wave and particle features of emitted light is sketched in Figure
\ref{fig1}. It shows two two-level atoms at a fixed distance $r$ that
are both coupled to the same free radiation field. In addition, the
atoms are excited by a resonant laser field, so that they can
continuously emit photons. Each emitted photon causes a ``click'' at a
certain point on a screen far away from the atoms. These ``clicks''
add up to an interference pattern with a spatial intensity
distribution very similar to a classical double-slit experiment.

This setup in Figure \ref{fig1} has been proposed by Scully and Dr\"uhl \cite{Scully82} in
1982 as a quantum eraser concerning observation and delayed choice
phenomena in quantum mechanics. Their predictions were verified
experimentally by Eichmann {\em et al.} in 1993 \cite{Eichmann}.
Since then the interpretation of this experiment has attracted
continuous interest (see for instance
Ref.~\cite{Englert,Itano,Schoen,Skornia} and references therein).
This is mainly due to its simplicity together with the fundamental
quantum mechanical issues it deals with.

Recently, Sch\"on and Beige \cite{Schoen} presented a quantum
mechanical description of the experiment based on the assumption of
environment-induced measurements. Here we first summarise the main
ideas of Ref.~\cite{Schoen} and then focus on the nature of the
spatial interference of the spontaneously emitted photons. Main
features of the interference pattern can also be predicted by a
corresponding classical model in which the atoms are replaced by
dipole sources of coherent light. Finally, similarities and
differences of the two descriptions are discussed.

\section{The spontaneous emission of photons}

The setup shown in Figure \ref{fig1} consists of two components ---
the quantum mechanical system including the atoms together with the
free radiation field and its environment represented by the screen. 
The time evolution of the quantum mechanical system can be described 
by the Schr\"odinger equation. Let us denote the energy
difference between the ground state $|1\rangle_i$ and the excited
state $|2 \rangle_i$ of atom $i$ at the position ${\bf r}_i$ by $\hbar
\omega_0$ whilst $a_{{\bf k}\lambda}^\dagger$ is the creation operator
of a single photon with wave vector ${\bf k}$ and polarisation
$\lambda$. The Hamiltonian that describes the interaction between the
atoms and the quantised free radiation field within the rotating wave
approximation equals
\begin{equation} \label{21}
H_{\rm int} = \hbar \sum_{i=1,2} \sum_{{\bf k}\lambda} {\rm e}^{
-{\rm i} {\bf k} \cdot {\bf r}_i} \, g_{{\bf k}\lambda} \, a_{{\bf
k}\lambda}^\dagger \, S_i^- + {\rm h.c.} ~,
\end{equation}
where $g_{{\bf k}\lambda}$ describes the coupling strength of the
field mode $({\bf k},\lambda)$ and $S_i^-$ is the atomic lowering
operator $|1 \rangle_i \langle 2|$.

The occurrence of spontaneous ``clicks'' at points on the screen
implies that the effect of the screen onto the quantum mechanical
system can be described with the projection postulate for ideal
measurements. The screen measures whether a photon has been created in
the free radiation field or not \cite{he}. If a photon is detected,
then the direction $\hat{\bf k}_0$ \cite{foot} of its wave vector
is determined \cite{Schoen}. The projector that describes this
measurement outcome is given by
\begin{equation} \label{pro} 
I\!\!P_{\hat{\bf k}_0} = \sum_{k,\lambda} |1_{k \hat{\bf k}_0 \lambda}
\rangle \langle 1_{k \hat {\bf k}_0 \lambda}| ~.
\end{equation}
If no ``click'' is observed, then the system is projected onto a state
with the free radiation field in the vacuum state $|0_{\rm ph}
\rangle$. Here the atoms are continuously driven by a laser field and
we assume that the screen performs rapidly repeated ideal
measurements. The time $\Delta t$ between two successive measurements
should not be too short, $\Delta t \gg 1/\omega_0$, to allow for a
substantial time evolution, but also not too big so that the
excitation of two-photon states in $\Delta t$ is negligible
\cite{he}. As these measurements are caused by the presence of the
screen we call them environment-induced measurements.

For simplicity, let us assume that the state of the atoms at a time
$t$ is known and equals $|\psi\rangle$ while the free radiation field
is in the ground state $|0_{\rm ph} \rangle$. Letting the system
evolve for a short time $\Delta t$ and applying the projector
(\ref{pro}) we find
\begin{equation} \label{eve}
|\psi\rangle |0_{\rm ph} \rangle \stackrel{\Delta t} {\longrightarrow}
\sum_{i=1}^2 \left[ \, {3 A \over 8 \pi} \, \big( \, 1 - | \hat{\bf
D}_{21} \cdot \hat{\bf k}_0 |^2 \, \big) \right]^{1/2} {\rm e}^{-{\rm
i} k_0 \hat{\bf k}_0 \cdot {\bf r}_i} \, S_i^-| \psi \rangle \, \left
| \begin{array}{c} {\sf normalised} \\[-.1cm] {\sf field~state} \,
(\hat{\bf k}_0) \end{array} \right \rangle ~,
\end{equation}
where the right hand side is the unnormalised state of the system in
case of a ``click'' in the $\hat{\bf k}_0$ direction away from the
atoms at time $t+\Delta t$. Here we assume that the dipole moment
$_{i} \langle 2|{\bf D}|1 \rangle_i = {\bf D}_{21}$ is for both atoms
the same and $A$ denotes the spontaneous decay rate of a single atom
in free space. Eq.~(\ref{eve}) has been derived with the help of first
order perturbation theory. For details see Section II in
Ref.~\cite{Schoen}.

The norm squared of the state (\ref{eve}) gives the probability
density $I_{\hat{\bf k}_0} (\psi)$ for the corresponding measurement
outcome,
\begin{equation} \label{I}
I_{\hat{\bf k}_0} (\psi) = {3 A \over 8 \pi} \, \big( \, 1 - |
\hat{\bf D}_{21} \cdot \hat{\bf k}_0 |^2 \, \big) \, \Big \| \, \sum_i
{\rm e}^{-{\rm i} k_0 \hat{\bf k}_0 \cdot {\bf r}_i} \, S_i^-| \psi
\rangle \, \Big \|^2 ~.
\end{equation}
The probability density for any ``click'' to occur can be obtained by
integrating over all orientations of $\hat{\bf k}_0$, leading to the
product of the spontaneous decay rate $A$ and the population of the
excited atomic states, namely $I(\psi) = A \sum_i \|S_i^- |\psi
\rangle \|^2$. The ``clicks'' on the screen are caused by the {\em
spontaneously emitted} photons. Immediately after the measurement, the
excitation in the free radiation field vanishes and its state becomes
again $|0_{\rm ph} \rangle$.

\section{The interference pattern on the screen}

\begin{center}
\begin{figure}[h]
\epsfig{file=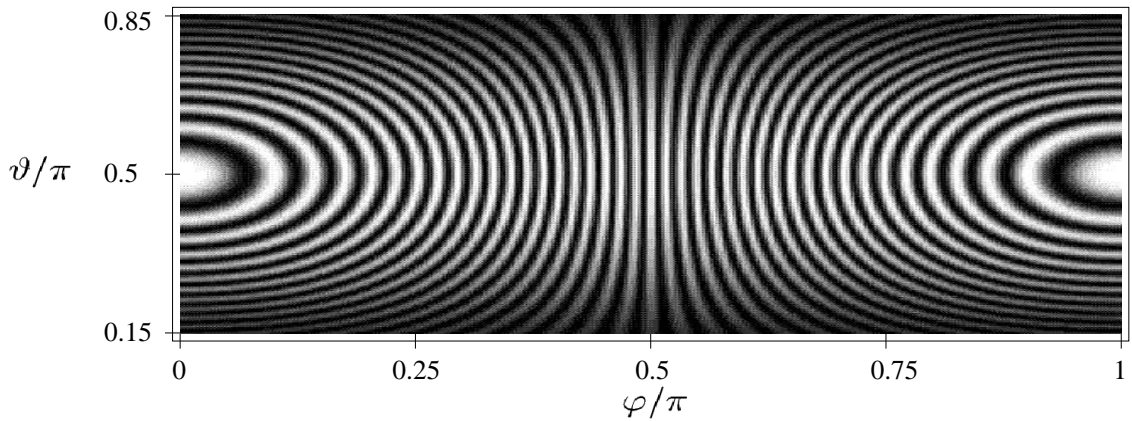,width=15cm} \\[-0.8cm]
\caption{ Density plot of the emission rate $I_{\hat{\bf
k}_0}(\rho_{\rm ss})$ for two continuously driven two-level atoms with
$r = 20 \, \lambda_0$ and $\Omega_1 = \Omega_2 = 0.3 \, A$. White
areas correspond to spatial angles $\vartheta$ and $\varphi$ with
maximal intensity.} \label{fig2}
\end{figure}
\end{center}

Let us now specify the spatial probability density for a photon
emission. If the atoms are continuously driven by a resonant laser
field, their state $|\psi\rangle$ shortly before an emission is not
known. To apply the result of Eq.~(\ref{I}) to this situation we 
describe the atoms by their steady state density matrix $\rho_{\rm
ss}$, so $I_{\hat{\bf k}_0}$ becomes
\begin{eqnarray} \label{5}
I_{\hat{\bf k}_0}(\rho_{\rm ss}) 
&=& \frac{3 A}{8 \pi} \, 
\big( \, 1 - | \hat{\bf D}_{21} \cdot \hat{\bf k}_0 |^2 \, \big) \,
\Big[ \, 2 \, \langle 22| \rho_{\rm ss} |22 \rangle
+ \langle 12| \rho_{\rm ss} |12 \rangle + \langle 21| \rho_{\rm ss} |21 \rangle \nonumber\\
& & \hspace*{4cm} + 2 \, {\rm Re} \, \Big( \langle 12| \rho_{\rm ss} |21 \rangle \,
{\rm e}^{-{\rm i} k_0 \hat{\bf k}_0 \cdot ({\bf r}_1 - {\bf r}_2)} \Big) \, \Big] ~.
\end{eqnarray}
In the following we denote the Rabi frequency of the laser field with
respect to atom $i$ by $\Omega_i$. Proceeding as in
Ref.~\cite{Schoen} to calculate $\rho_{\rm ss}$ we find that
\begin{eqnarray} \label{Iss}
I_{\hat{\bf k}_0}(\rho_{\rm ss}) &=& \frac{3 A}{8 \pi} \, {1 - |
\hat{\bf D}_{21} \cdot \hat{\bf k}_0 |^2 \over (A^2 + 2
|\Omega_1|^2)(A^2 + 2 |\Omega_2|^2)} \, \Big[ \, 4 \, |\Omega_1|^2
|\Omega_2|^2 + A^2|\Omega_1|^2 + A^2|\Omega_2|^2 \nonumber \\ &&
\hspace*{5.8cm} + 2 A^2 \, {\rm Re} \, \Big( \Omega_1^* \Omega_2 \,
{\rm e}^{-{\rm i} k_0 \hat{\bf k}_0 \cdot ({\bf r}_1 - {\bf r}_2)}
\Big) \, \Big] ~.
\end{eqnarray}
Figure \ref{fig2} shows the spatial intensity distribution for the
case where both laser fields are in phase and have the same intensity,
i.e.~$\Omega_1=\Omega_2$. One easily recognises an interference
pattern which results from the last term in Eq.~(\ref{Iss}).

To understand the origin of the interference let us assume again that
the state of the atoms shortly before an emission is known and equals
$|\psi \rangle$ while the free radiation field is in the vacuum
state. During the time evolution $\Delta t$ with respect to the
Hamiltonian (\ref{21}) each atom transfers excitation into all modes
$({\bf k},\lambda)$ of the free radiation field. In case of a photon
detection the state of the field is projected onto a photon state with
the wave vector towards the direction $\hat{\bf k}_0$ of the
``click''. To calculate the probability density for this event, one
had to determine the norm of the reset state given in
Eq.~(\ref{eve}). This state is the sum of the contributions from both
atoms which differ only by the phase factor ${\rm e}^{-{\rm i} k_0
\hat{\bf k}_0 \cdot ({\bf r}_1-{\bf r}_2)}$. In addition, each
contribution contains a different atomic state, namely $S_1^- |\psi
\rangle$ or $S_2^- |\psi \rangle$. Therefore the visibility of the
interference pattern depends on the overlap of the atomic states
$S_1^- |\psi \rangle$ and $S_2^- |\psi \rangle$. If these two states
are orthogonal, i.e.~when the atoms are prepared in $|22\rangle$, then
the probability density to find a ``click'' at a certain point on the
screen is the sum of the probabilities of two one-atom
cases. Maximum interference takes place if $S_1^- |\psi \rangle=S_2^-
|\psi \rangle$. This is the case when $|\psi\rangle$ is a non-trivial
superposition of the states $|12 \rangle$ and $|21 \rangle$. 

\noindent

\vspace*{-0.5cm} In general both atoms contribute to the spontaneous
emission of a photon. Nevertheless, one could ignore this and ask from
which atom the photon originated. To answer this one has to perform a
measurement on the atomic state and determine whether the atoms are
either in $S_1^- |\psi \rangle$ or $S_2^- |\psi \rangle$. These two
states are the reset states of the two one-atom cases where one atom
emits and the state of the other one remains unchanged. Only if the
reset states are orthogonal, one can find out with certainty which
atom emitted the photon and the {\em which way} information is
available in the experiment. In this case the interference
vanishes. The more overlap the states have, the stronger becomes the
visibility of the interference pattern
\cite{Scully82,Eichmann,Itano}. For a more detailed discussion of
interference criteria see also Refs.~\cite{Englert,Schoen,Skornia} and
references therein.

\section{Comparison with the classical double-slit experiment}

Let us now consider a classical double-slit experiment with two dipole
sources at positions ${\bf r}_1$ and ${\bf r}_2$. Both dipoles have
the same direction $\hat{\bf D}$ of the dipole moment and
simultaneously emit electromagnetic waves with frequency
$\omega_0$. The resulting electric field at a certain point ${\bf R}$
on a far away screen ($|{\bf R} -{\bf r}_i| \gg |{\bf r}_1-{\bf
r}_2|$) is then
\begin{equation}
{\bf E}({\bf R},t) = \sum_{i=1}^2 {E_0^{(i)} \over |{\bf R} -{\bf
r}_i|} \, \Big[ \, \hat{\bf D} - (\hat{\bf k}_0 \cdot \hat{\bf D})
\hat{\bf k}_0 \, \Big] \, {\rm e}^{-{\rm i} {\bf k}_0 \cdot ({\bf R-
r}_i)} \, {\rm e}^{-{\rm i} \omega_0 t} \, ~,
\end{equation}
where $E_0^{(1)}$ and $E_0^{(2)}$ characterise the strength of the
dipoles. The phase of the electric field produced by each source $i$
is sensitive to its position ${\bf r}_i$ while the wave vector ${\bf
k}_0 = k_0 \hat{\bf k}_0$ has in both cases, to a very good
approximation, the same direction as ${\bf R} - {\bf r}_1 \approx {\bf
R} - {\bf r}_2$ and the amplitude $k_0$. The intensity of the produced
light is given by
\begin{equation} \label{last}
I_{\hat{\bf k}_0} (E_0^{(1)},E_0^{(2)}) = {\epsilon_0 c \over 2} \,
\big( \, 1 - | \hat{\bf D} \cdot \hat{\bf k}_0 |^2 \, \big) \, \Big[
\, \big|E_0^{(1)} \big|^2 + \big|E_0^{(2)} \big|^2 + 2 \, {\rm Re} \,
\Big( E_0^{(1)*} E_0^{(2)} \, {\rm e} ^{-{\rm i} {\bf k}_0 \cdot ({\bf
r}_1-{\bf r}_2)} \Big) \, \Big] ~.
\end{equation}
The last term describes the interference pattern on the far away
screen.

If we compare this spatial dependence of the interference with the
spatial dependence in the quantum mechanical double-slit experiment,
we see that they can in both cases be exactly the same. Nevertheless
the visibility in the quantum model is in general lower than in the
classical model. The reason is that population of the atomic state
$|22\rangle$ does not contribute to the amplitude of the interference
term. In the quantum case, the more {\em which way} information is
available in the experiment, the less visibility is found in the
interference pattern \cite{Englert}. This reduction of the visibility
is a purely quantum mechanical effect.

The analogies of the quantum system with an equivalent classical one
is not a surprising feature \cite{Jiannis}. Each atom populates the
free radiation field with an excitation of a certain effective
frequency, $\omega_0$, as dictated from the energy difference between
the atomic levels $1$ and $2$. In addition, when the measurement of
the radiation field takes place and a photon is detected then the
state of the field is projected towards a certain direction $\hat {\bf
k}_0$, as can be seen from Eq.~(\ref{eve}). These two deterministic
characteristics are not imposed in the initial theory as the atoms are
allowed to emit with any frequency and the radiation field around them
has no preference in the direction of propagation of its photon
states. Nevertheless, the average along all possible photon
frequencies as well as the act of detection, which projects the free
radiation field onto the photon state with direction $\hat {\bf k}_0$,
gives the equivalence between the spontaneously emitting
atoms and the classical sources.

\section{Conclusions}

Here we presented a quantum mechanical description of a two-atom
double-slit experiment. We showed that spontaneous emission of photons
can be derived from basic quantum mechanical principles by assuming
environment-induced measurements on the free radiation field. By
calculating the probability density for an emission into a certain
direction $\hat{\bf k}_0$ we found spatial interference. The
interference fringes appearing in the quantum case result from the
spontaneously emitted photons and are not just produced by the
interference of a scattered laser field. This is particularly evident
in Eq.~(\ref{I}) where the atoms are initially prepared in an atomic
state $|\psi\rangle$ without the need for this state to be prepared by
continuous laser radiation.

In general, the spatial dependence of the interference pattern and its
visibility could also be described by an equivalent classical
double-slit experiment. The superposition
of the two amplitudes in the quantum and in the classical case
produces standing waves throughout space. Their pattern is observed on
the screen as seen in Figure \ref{fig2}.


\begin{thebibliography}{77}
\bibitem{foot} 
In the following we mark normalised vectors with a hat.
\end{thebibliography}
\end{document}